\documentclass[aj,twocolumn]{aastex62}
\pdfoutput=1 
\usepackage{amsmath,amstext}
\usepackage[T1]{fontenc}
\usepackage[utf8]{inputenc}
\usepackage{apjfonts} 
\usepackage[figure,figure*]{hypcap}
\usepackage{amsmath,amssymb,amsfonts,graphicx}
\usepackage{todonotes}
\usepackage{color}
\usepackage{hyperref}
\usepackage{cleveref}
\usepackage{graphicx}
\usepackage{tabularx}
\usepackage{footnote}
\usepackage{subfigure}

\usepackage{todonotes}   

\shorttitle{Circumnuclear rings and Lindblad resonances}
\shortauthors{Schmidt et al.}

\received{April 10, 2019}
\accepted{June 8, 2019}

\submitjournal{AJ}

\begin{document}

   \title{Circumnuclear rings and Lindblad resonances in spiral galaxies}

\author{Schmidt, E. O.}
\affiliation{Universidad Nacional C\'ordoba. Observatorio Astron\'omico de C\'ordoba. C\'ordoba, Argentina.}
\affiliation{Instituto de Astronomía Te\'orica y Experimental (IATE). C\'ordoba, Argentina.}
\affiliation{Consejo de Investigaciones Cient\'{i}ficas y T\'ecnicas de la Rep\'ublica Argentina, Avda. Rivadavia 1917, C1033AAJ, CABA, Argentina.}
\email{eduschmidt@oac.unc.edu.ar}

\author{Mast D.\thanks{ddd}}
\affiliation{Universidad Nacional C\'ordoba. Observatorio Astron\'omico de C\'ordoba. C\'ordoba, Argentina.}
\affiliation{Consejo de Investigaciones Cient\'{i}ficas y T\'ecnicas de la Rep\'ublica Argentina, Avda. Rivadavia 1917, C1033AAJ, CABA, Argentina.}
\affiliation{Visiting Astronomer at Complejo Astron\'omico El Leoncito (CASLEO) operated under agreement between Consejo Nacional de Investigaciones Cient\'{i}ficas y T\'ecnicas de la Rep\'ublica Argentina and the National Universities of La Plata, C\'ordoba, and San Juan.}

\author{D\'{i}az R.J.}
\affiliation{Universidad Nacional C\'ordoba. Observatorio Astron\'omico de C\'ordoba. C\'ordoba, Argentina.}
\affiliation{Consejo de Investigaciones Cient\'{i}ficas y T\'ecnicas de la Rep\'ublica Argentina, Avda. Rivadavia 1917, C1033AAJ, CABA, Argentina.}
\affiliation{Gemini Observatory, 950 N Cherry Ave., Tucson, AZ85719, USA.}

\author{Ag\"{u}ero M.P.}
\affiliation{Universidad Nacional C\'ordoba. Observatorio Astron\'omico de C\'ordoba. C\'ordoba, Argentina.}
\affiliation{Consejo de Investigaciones Cient\'{i}ficas y T\'ecnicas de la Rep\'ublica Argentina, Avda. Rivadavia 1917, C1033AAJ, CABA, Argentina.}

\author{G\"{u}nthardt G.}
\affiliation{Universidad Nacional C\'ordoba. Observatorio Astron\'omico de C\'ordoba. C\'ordoba, Argentina.}

\author{Gimeno, G.}
\affiliation{Gemini Observatory, 950 N Cherry Ave., Tucson, AZ85719, USA.}

\author{Oio, G.}
\affiliation{Universidad Nacional C\'ordoba. Observatorio Astron\'omico de C\'ordoba. C\'ordoba, Argentina.}
\affiliation{Instituto de Astronomía Te\'orica y Experimental (IATE). C\'ordoba, Argentina.}
\affiliation{Consejo de Investigaciones Cient\'{i}ficas y T\'ecnicas de la Rep\'ublica Argentina, Avda. Rivadavia 1917, C1033AAJ, CABA, Argentina.}

\author{Gaspar G.}
\affiliation{Universidad Nacional C\'ordoba. Observatorio Astron\'omico de C\'ordoba. C\'ordoba, Argentina.}
\affiliation{Consejo de Investigaciones Cient\'{i}ficas y T\'ecnicas de la Rep\'ublica Argentina, Avda. Rivadavia 1917, C1033AAJ, CABA, Argentina.}

\begin{abstract}

   {}
   {In order to study the location of circumnuclear rings (CNR) and their possible relation with the inner Lindblad resonances (ILR), we investigate a sample of spiral galaxies. For this purpose, we have obtained and analyzed medium resolution spectra of 5 spiral galaxies in the range 6200 \AA \ to 6900 \AA.\, Through the H$\alpha$ emission line, we constructed the radial velocity curves, and then the rotation curves. By fitting them, considering two or three components of an axisymetric Miyamoto$-$Nagai gravitational potential, we constructed the angular velocity and Lindblad curves. In addition, we determined the CNR radius by using the 2D spectra and generating the H$\alpha$ spatial emission radial profiles.\\}
   {We determined the position of the resonances and we calculated the angular velocity pattern, which are in the range of 26 $-$ 47 km s$^{-1}$ kpc$^{-1}$ for the galaxies of the sample. According to our results, the CNRs are located between the inner ILR (iILR) and the outer ILR (oILR), or between the center of the galaxy and the ILR, when the object has only one of such resonance; in agreement with previous results. In addition, we calculated the dimensionless parameter defined as $\mathcal{R}=$ R$_{CR}$ / R$_{bar}$, being in the range 1.1 $-$ 1.6, in agreement with previous results found in the literature. }
 

\end{abstract}


\section{Introduction} \label{sec:intro}

The scientific community widely accepts the accretion of gas to the central super massive black hole as the cause of nuclear activity in galaxies. These black holes would be a common component in most galaxies with a sufficiently massive bulge \citep[e.g.,][]{Ferrarese2005}. However, only half (about 43$\%$) of the Local Universe galaxies host an active galactic nuclei (AGN), as was suggested by \cite{Ho1997}. This discrepancy between the presence of the super massive black holes and nuclear activity must be studied in relation to the possibility and efficiency of transporting gas to the central region of the galaxy.

Circumnuclear rings (CNR) have radii ranging from 50 pc to 2 kpc and would be in the central regions of almost 20$\%$ of disk galaxies in the Local Universe with a local density of $0.54\pm0.12$ galaxies/Mpc$^{3}$ \citep{Diaz2004,Knapen2005,Comeron2010}. CNRs act as repositories of gas, slowing the flow of gas towards the center, and have a strong star formation, reaching in some cases more than 50 times the star formation rate of a galactic disk \citep{Kennicutt1998}. Depending on the duration of the ring, they may be responsible for the creation of an important stellar mass in the central regions \citep{VanDerLaan2012}.

The increase of nuclear activity in disk galaxies, whether in the form of Starburst or AGN, seems to be closely linked to the evolution of the galaxy. It has been observed that both forms of activity co-exist \citep[e.g.,][]{Heckman1997} and are a clear manifestation of the symbiotic evolution of the galactic centers and their host galaxies. The close correlation observed between central black holes masses and the velocity dispersion of stars in bulges \citep{Ferrarese2005} provides direct evidence of this evolution and allows to study the structure, dynamics, and evolution of the galaxies. At the same time, it is found that the kinematics of the gas clouds in the nucleus of the AGN is strongly influenced by the presence of the central black hole \citep[e.g.,][]{Merritt2001,Tremaine2002,Schmidt2016,Schmidt2018,Schmidt2019}.

To boost and fuel the nuclear activity in the AGN or the nuclear Starburst, there must be efficient gas conduction to the core. This process must be accompanied by a substantial loss of angular momentum of the gaseous material by almost 8 orders of magnitude. This suggests the existence of gravitational torques that act through bars or interactions of galaxies. The non-axysimmetric nature of the mass distribution facilitates the loss of angular momentum in the infalling material \citep[e.g.,][]{Shlosman1990, Combes2001}.

There are many doubts about what are the proper conditions of temperature, density, and dynamics that trigger nuclear activity, and what are the parameters that regulate the duration of these outbreaks. For example, in the galaxy NGC 1241 the presence of several perturbations was detected in scales of hundreds of parsecs linked to the CNR \citep{Diaz2003}. Some of these disturbances are good candidates to remove the angular momentum of the gas and feed the Seyfert 2 nucleus of this galaxy. This motivated the systematic search for the presence of nuclear activity in galaxies with CNR, which suggested that the appearance of rings in the central regions seems to be related to the intrinsic properties of host galaxies, with self-instabilities, minor fusions or the capture of giant molecular clouds \citep[e.g.,][]{Corbin2000, Mast2006, Diaz2006}. There is evidence that points to the CNR as tracers of the feeding of nuclear activity. \citet{Arsenault1989} reported a higher incidence of the combination of bars and rings in Starburst and AGN host galaxies, with respect to normal galaxies. On the other hand, \citet{Hunt1999} found that Seyfert galaxies have more rings (interior and exterior) than normal galaxies. The study by \citet{Aguero2016} suggests a simultaneity between Seyfert activity and CNR greater than that expected for the morphological type distribution of host galaxies. The ratio between Seyfert 2 / Seyfert 1 with CNR would be, according to these authors, 3 times greater than the expected considering galaxies without rings, in opposition to the ratio that predicts the geometric paradigm of the classic unified model for AGN. 

The preferred position of the rings with respect to the resonances in the disk of the galaxies is an open problem, which also motivates our work. According to the most accepted theory, the CNR are formed close to the internal Lindblad resonances (ILR, \citealt{Lindblad1964}) when the gas loses angular momentum and accumulates in that region \citep[e.g.,][]{Shlosman1990,Knapen1995,Combes1996, Comeron2010, Pf2014}. It was suggested that CNRs are formed between the center and the ILR, when there is only one of such resonance, or between the iILR and the oILR when there are two resonances \citep[e.g.,][]{Combes1996}. In this context, CNRs form because of resonant interactions of the gas with the inner resonances. Related to this, the bar torque changes its sign when it crosses each ILR. For example, in the case of only one ILR, the gas inside (outside) the resonance receives a positive (negative) torque and then it moves outward (inward). \\

On the contrary, some authors argue that ring formation is not due to resonances but by the centrifugal barrier that the inflowing gas along dust lanes cannot overcome \citep[e.g.,][]{WoongKim,Kim2012, Li2015, Seo2014}. In this scenario, the formation and location of the rings are not directly determined by the ILR and there would be no direct connection between them \citep[e.g.,][]{Kim2012, Ma2018}. From hydrodynamic numerical simulations it is shown that the presence of x2 orbits is the requirement for the formation of CNR and that there are no internal Lindblad resonances in barred galaxies \citep{Regan2003}. \citet{Li2015} suggest that knowing the radius of the ILR would be insufficient to find the location of the CNR. 


\citet{Comeron2010} studied the sizes and shapes of CNRs in a sample of more than 100 galaxies and they found that objects with stronger bars host smaller rings. According to them, the ring ellipticity is in the range of 0 $-$ 0.4. It was reported that the size of the rings is well correlated with the galaxy mass distribution \citep{Mazzuca2011}. According to these authors, a higher mass concentration results in a smaller ring. They also found that the CNR size is dependent on the bar strength, in agreement with recent results of \citet{Seo2019}.

In the present work we study a sample of 5 galaxies with CNR reported in the literature using long-slit optical spectra, to provide elements that allow us to shed light on the dynamic nexus between CNR and nuclear feeding mechanisms. In particular we studied the CNR position and their relation with Lindblad resonances. This paper is structured as follows: Sec.\ref{sec:sample} introduces the observed sample and briefly describes the observations, Sec.\ref{sec:analysis} presents the complete kinematical analysis, and Sec.\ref{sec:discussion} offers a discussion of the obtained results. Finally, Sec.\ref{sec:summary} provides a brief summary of the work and the conclusions.


\section{Sample and observations}\label{sec:sample}

For the present work, we selected a sub-sample of 5 spiral galaxies from a mother sample of nearby galaxies with CNR, observable from the southern hemisphere and with spatial scales that allow us to resolve the rings with 2-4 m class telescopes. Table \ref{tab:sample} lists the main properties that characterize the galaxies, such as the object name, right ascension, declination, distance, scale, and apparent magnitude in the $R$ filter. The data were taken from the Nasa Extragalactic Database (NED)\footnote{The NASA/IPAC Extragalactic Database (NED)
is operated by the Jet Propulsion Laboratory, California Institute of Technology,
under contract with the National Aeronautics and Space Administration.}. 

Observations were performed in different campaigns between 2002 and 2004 using the REOSC spectrograph at the 2.15 m telescope of the Complejo AStron\'omico el LEOncito (CASLEO), in Argentina. All spectra were taken in the long-slit mode. The detector is a Tektronix 1024 $\times$ 1024 CCD with 24 $\mu$m pixels. We used a 1200 line mm$^{-1}$ grating giving a spectral resolution of $\sim$ 3500 in the range 6200 \AA \ to 6900 \AA. The dispersion was 32 \AA /mm$^{-1}$, the reciprocal dispersion 0.76 \AA \ pixel$^{-1}$, the angular scale 1\farcs02 pixel$^{-1}$ and  the spectral resolution 1.9 \AA. The average seeing during the observations was $\sim2\farcs5$. The standard procedure was followed for the observations and data reduction, including correction for bias, dark current, and flat field. The spectra were wavelength calibrated using comparison lines from a Cu-Ne-Ar lamp and three standard stars were observed each night to flux calibrate the spectra. Data reduction was carried out using IRAF\footnote{IRAF is distributed by the National Optical Astronomy
Observatories, which is operated by the Association of Universities for
Research in Astronomy, Inc. (AURA) under cooperative agreement with the National
Science Foundation} reduction package.  Additional observational information can be found in \cite{Mast2005}. 

\begin{table*}[!ht]
\caption{Galaxy sample}             
\label{tab:sample}      
\centering          
\begin{tabular}{l c c c c c}    
\hline\hline     
Galaxy & R.A. & Decl. & D (Mpc) & Scale (kpc arcsec$^{-1}$) & mag. \\
(1) & (2) & (3) & (4) & (5) & (6) \\
\hline

IC 1438 & 22:16:29.09 & -21:25:50.5 & 34.7 & 0.168 & 11.57(*) \\ 
IC 4214 & 13:17:42.69 & -32:06:06.1 & 29.6 & 0.144 & 10.72(*)  \\
NGC 1512 & 04:03:54.28 & -43:20:55.9 & 10.2 & 0.050 & 11.13  \\
NGC 2935 & 09:36:44.85 & -21:07:41.3 & 28.3 & 0.137 & 11.92  \\
NGC 6753 & 19:11:23.64 & -57:02:58.4 & 42.5 & 0.206 & 11.74 \\
\hline                  
\end{tabular}
\footnote{Column (1): target ID. Column (2): R.A. (J2000). Column (3): Decl. (J2000). Column (4): Distance in Mpc. Column (5): Scale in kpc arcsec$^{-1}$. Column (6): R-band calibrated magnitude \citep{2005MNRAS.361...34D}, (*) R (Cousins) (R$_{T}$) magnitude \citep{1989spce.book.....L}.}
\end{table*}

\section{Analysis}\label{sec:analysis}

\subsection{Radial velocity curves}\label{sec:radial}

\begin{figure}[ht!]
\begin{center}
\includegraphics[clip,width=\columnwidth]{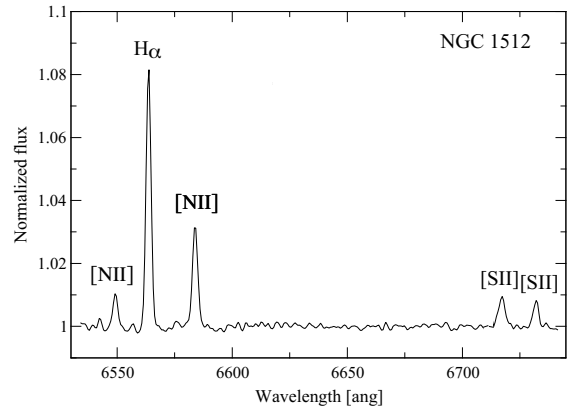}
\caption{Restframe NGC 1512 sample spectrum with the main emission lines indicated.}
\label{fig:spectra}
\end{center}
\end{figure}

Once the spectra were reduced, we proceeded to make extractions using one pixel wide apertures along the slit, covering the entire  spatial range with H$\alpha$ emission. Figure \ref{fig:spectra} shows one of these extractions, as an example, for the galaxy NGC 1512. This spectrum depicts the more conspicuous emission lines at this wavelength range, i.e. H$\alpha\lambda$6563 \AA, [NII]$\lambda\lambda$6548,6584 \AA, and [SII]$\lambda\lambda$6713,6731 \AA. 

\begin{table*}[!ht]
\caption{Radial velocity curves properties}             
\label{tab:radvelcur}      
\centering          
\begin{tabular}{l c c c c}    
\hline\hline     
Galaxy & Central gradient & Vr max. & r$_{Vrmax}$ \\
  & [km sec$^{-1}$ arcsec$^{-1}$] & [km sec$^{-1}$] &  [arcsec] \\
(1) & (2) & (3) & (4) \\
\hline

IC 1438 & 17 & 83 & 5 \\ 
IC 4214 & 24 & 210 & 9  \\ 
NGC 1512 & 9 & 140 & 13  \\ 
NGC 2935 & 4 & 90 & 45  \\ 
NGC 6753 & 17 & 193 & 12 \\ 
\hline                  
\end{tabular}
\footnote{Column (1): target ID. Column (2): radial velocity gradient in the inner part of the curve, determined between both velocity maxima. Column (3): maximum velocity reached by the curve. Column (4): position where the maximum is reached.}
\end{table*}

\begin{figure*}[ht!]
\begin{center}
\includegraphics[clip,trim=115 70 60 60,height=0.93\textheight,keepaspectratio=no]{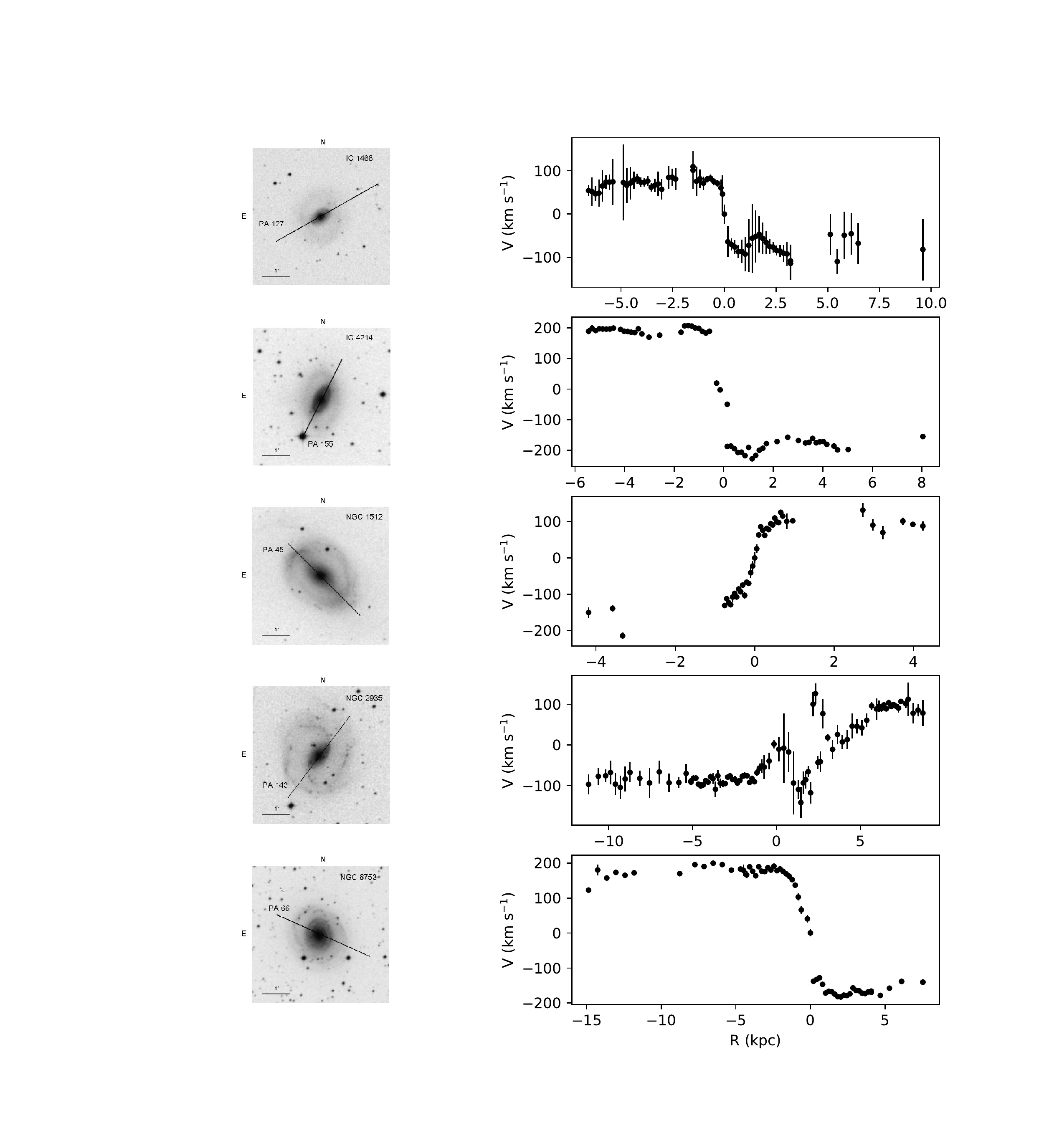}
\caption{\textit{Left}: galaxy images with the slit positions indicated. \textit{Right}: H$\alpha$ radial velocity curves of the observed galaxies. The error bars were determined using the formula of \citet{Keel1996}. Note that in some cases the size of the error bar is smaller than the size of the used symbol.}
\label{fig:radialvelocity}
\end{center}
\end{figure*}

Using the IRAF task \textit{splot}, we fitted gaussians to each emission line to determine its parameters: intensity, FWHM, flux, and centroid. In the present work we will focus on the kinematical study derived from H$\alpha$. Once radial velocities of each extracted spectrum have been determined, we constructed the radial velocity curves for each galaxy. On Table \ref{tab:radvelcur} we summarized the main properties of the curves. Images of the galaxies, the location of the slit, and the radial velocity curves are shown in Figure \ref{fig:radialvelocity}. The adopted systemic radial velocity (V$_{syst}$) for each galaxy are 2621 km s$^{-1}$ for IC 1438, 2330 km s$^{-1}$ for IC 4214, 924 km s$^{-1}$ for NGC 1512, 2271 km s$^{-1}$ for NGC 2935, and 3160 km s$^{-1}$ for NGC 6753. These velocities are the ones that better symmetrizes the radial velocity curves and they are in agreement with values found in the literature. For IC 4214 and NGC 2935, V$_{syst}$ does not coincide with V$_{obs}$ at the continuum peak, so an offset has to be applied to symmetrize the radial velocity curves. This offset, nevertheless, is always a fraction of the resolution element ($\sim$ 1 arcsec) and we do not have enough spatial information to make conjectures about possible off-centered nucleus, although we do not discard this possibility. The uncertainties in the radial velocities were calculated using the formula presented in \citet{Keel1996}.

All radial velocity curves are relatively smooth, with the exception of NGC 2935 which deviates from circular rotation around the galaxy center from $\sim1$ to 3 kpc on the red side of the curve.

\subsection{Rotation curves and resonances}
\label{sec:resonances}

In order to determine the rotation curves of the galaxies, we calculated the inclination angle of each object. Through isophotal fitting to Digitized Sky Survey (DSS)\footnote{The Online Digitized Sky Surveys server at the ESO Archive} $R$-band images, we measured the major and minor axis of each galaxy and we calculated the inclination angle considering that cos $i=\mathcal{B}/\mathcal{A}$, where  $\mathcal{B}$ and  $\mathcal{A}$ are the minor and major axis, respectively, of the outer isophote. We assumed the photometric position angle (PA$_{phot}$) as the angle measured from north to east of the outer isophote major axis.   Once obtained the inclination angle, and assuming that PA$_{phot}$ coincides with the kinematic PA (PA$_{kin}$), the circular rotation velocity can be calculated. After a small correction due to the difference between PA$_{obs}$ (position angle of the observation) and PA$_{kin}$, we are able to determine the circular velocity by:
\begin{equation}
\label{eq:vel}
V_{c}= \frac{V_{obs}-V_{syst}}{sin(i)}
  \end{equation}

where V$_{c}$ is the circular velocity, V$_{obs}$ is the observed radial velocity and V$_{syst}$ is the systemic radial velocity. While, in general, PA$_{kin}$ is defined as the mean value of the radial evolution of the kinematic PA, our adopted PA$_{phot}\sim$ PA$_{kin}$ was consistent with the values found in the literature for our objects. 

We fitted the rotation curves of the galaxies considering two or three mass components, depending on the case, corresponding to axisymetric Miyamoto$-$Nagai gravitational potentials \citep{Miyamoto-Nagai1975}:
\begin{equation}
\label{eq:pot}
    \Phi (R,z) = - \frac{G M}{\sqrt{R^2 + [a + \sqrt{z^2 + b^2}}]^2} 
    \end{equation}

where $\Phi (R,z)$ is the Miyamoto$-$Nagai potential at $(R, z)$ position, $M$ is the total mass, and $a$ and $b$ are shape parameters. The ratio b/a $\sim$ 0.4 corresponds to a flattened disk distribution, b/a $=$ 1 to an ellipsoidal distribution and b/a $\sim$ 5 to a spherical distribution  \citep[e.g.,][]{Binney}.

 The galaxies IC 4214, NGC 1512, NGC 2935 and NGC 6753 were fitted considering the presence of only two galactic subsystems (a central ellipsoidal component and a flattened disk) without the need for the external spherical component (halo). On the other hand, IC 1438 rotation curve was fitted considering the presence of three galactic subsystems: a central ellipsoidal component, a flattened disk, and a massive halo.      

Table \ref{tab:params} lists the obtained parameters that optimize the $\chi^{2}$ minimization fit, such as the scale parameters and the total mass. 

\begin{table*}[!ht]
\caption{Rotation curves parameters considering the fitted Miyamoto$-$Nagai model.}             
\label{tab:params}      
\centering          
\begin{tabular}{c c c c}     
\hline\hline       
Galaxy & Component 1 & Component 2 & Component 3   \\
\hline                    
   IC1438 & a$_{1}=$ 0.20 kpc                              & a$_{2}=$ 1.50 kpc             & a$_{3}=$ 7.0 kpc     \\      
          & b$_{1}$/a$_{1} =$1                             & b$_{2}$/a$_{2} =$0.33         & b$_{3}$/a$_{3} =$1.43   \\  
          & M$_{1}=$ 5.6 $\times$ 10$^{9}$ M$_{\odot}$     & M$_{2}=$ 2.0 $\times$ 10$^{10}$ M$_{\odot}$  & M$_{3}=$ 2.1 $\times$ 10$^{11}$ M$_{\odot}$   \\
 \hline  
  IC4214 & a$_{1}=$ 0.25 kpc                               & a$_{2}=$ 3.14 kpc     &   \\   
         & b$_{1}$/a$_{1} =$ 0.92                          & b$_{2}$/a$_{2} =$0.63     &  \\  
         & M$_{1}=$ 2.1 $\times$ 10$^{10}$ M$_{\odot}$     & M$_{2}=$ 1.4 $\times$ 10$^{11}$ M$_{\odot}$   &  \\
 \hline         
  NGC 1512  & a$_{1}=$ 0.20 kpc                               & a$_{2}=$ 2.5 kpc           & \\
            & b$_{1}$/a$_{1} =$ 1                          & b$_{2}$/a$_{2} =$0.20       & \\
            & M$_{1}=$ 3.4 $\times$ 10$^{9}$ M$_{\odot}$     & M$_{2}=$ 2.7 $\times$ 10$^{11}$ M$_{\odot}$   & \\    
   \hline          
 NGC 2935   & a$_{1}=$ 0.50 kpc                               & a$_{2}=$ 2.8 kpc           & \\
            & b$_{1}$/a$_{1} =$ 1                          & b$_{2}$/a$_{2} =$0.61         & \\
            & M$_{1}=$ 6.8 $\times$ 10$^{9}$ M$_{\odot}$     & M$_{2}=$ 3.6 $\times$ 10$^{10}$ M$_{\odot}$     & \\   
  \hline           
NGC 6753    & a$_{1}=$ 0.68 kpc                               & a$_{2}=$ 4.7 kpc           & \\
            & b$_{1}$/a$_{1} =$ 0.73                          & b$_{2}$/a$_{2} =$0.49      & \\
            & M$_{1}=$ 7.9 $\times$ 10$^{10}$ M$_{\odot}$     & M$_{2}=$ 3.4 $\times$ 10$^{11}$ M$_{\odot}$     & \\               
            
\hline                  
\end{tabular}
\end{table*}

Considering the potential, the circular velocity can be inferred as V$_{c}=\sqrt{-R (\partial \Phi / \partial R)}$, \citep[see for example][]{Binney}. Once obtained the circular velocity, the angular velocity can be easily calculated as $\Omega =$ V$_{c}$/R and the epicyclic frequency can be inferred \citep[e.g.,][]{Elmegreen}:
\begin{equation}
\label{eq:frecuency}
    \kappa^2 = 4 \Omega^2 [1 + \frac{1}{2}(\frac{R}{\Omega} \frac{d\Omega}{dR})]
    \end{equation}
    
The location of the ILR and OLR can be calculated taking into account that $\Omega - \Omega_{p} = - \kappa$/2 at OLR and $\Omega - \Omega_{p} = \kappa$/2 at ILR \citep[e.g.,][]{Binney,Elmegreen}, where $\Omega_{p}$ is the pattern speed. In addition, the location of the corotation resonance (CR) can be inferred when $\Omega = \Omega_{p}$.      

In order to construct the Linblad $\Omega - \kappa$/2 and $\Omega + \kappa$/2 curves for the galaxies, the first thing we did was to determine the position of the CR. There is more than one way to calculate the radius of this resonance. For example, in order to do this \citet{Puerari_Dottori97} presented a method based on the Fourier analysis of azimuthal profiles \citep{Aguerri1998, Diaz2003}. On the other hand, \cite{Prendergast1983} suggested that the dust lanes indicate the location of the shock regions in the spiral arms. This interpretation was first quantified by \citet{Roberts1979}. Inside the corotation radius, dust lanes are located in the trailing side of the bar or the spiral arms perturbation pattern. In this locations the gas in circular orbits in the disk encounters the perturbation pattern, which is moving at lower angular velocity.  Outside the corotation radius it is expected that the dust lanes are in the leading side of the perturbation pattern. The radii in which the dust lanes change from the trailing to the leading side of the perturbation pattern, indicate the approximate location of the CR. Taking into account all these considerations, we determined the CR radius by measuring directly from DSS images.



Considering the measured length of the bar (R$_{bar}$) and the corotation radius (R$_{CR}$), the dimensionless parameter $\mathcal{R}$, defined as $\mathcal{R}=$ R$_{CR}$ / R$_{bar}$, is usually useful to characterize the bars as fast or slow. For the galaxies in our sample we obtained values for $\mathcal{R}$ in the range 1.1 $-$ 1.6, in agreement with previous results \citep[e.g.,][]{Debattista2000, Rautiainen2008, Aguerri2015, Guo2019}.

The rotation curves corresponding to the fitted axial symmetric model are illustrated in the left panels of Figure \ref{fig:curves}, as well as the angular velocity curves and the Lindblad $\Omega - \kappa$/2 and $\Omega + \kappa$/2 curves (right panel) of the studied galaxies. Note that in all left panels of this figure, V$_{circ}$ is 0 near the center. This is mostly likely due to limited resolution, taking into account that if there is a central mass concentration (e.g., a black hole), the rotation velocity should rise toward the center. The net effect of the finite spatial resolution is to average plus and minus values in both sides of nucleus along the major axis \citep{Sofue2013}. The spectra have average spatial resolutions of 2\farcs5, with a 1\farcs02 spatial sampling. The innermost resolved rotation velocities are of the order of tens of km s$^{-1}$ at radii of 100-300 pc, without a change of slope towards the galactic core. This means the gravitational sphere of influence of the supermassive black hole is not resolved in the rotation curve for the average distance of the sample.

\begin{figure*}[ht!]
\centering
\includegraphics[angle=0,trim = 15mm 20mm 20mm 25mm,clip, width=\textwidth,height=0.899\textheight,keepaspectratio=no]{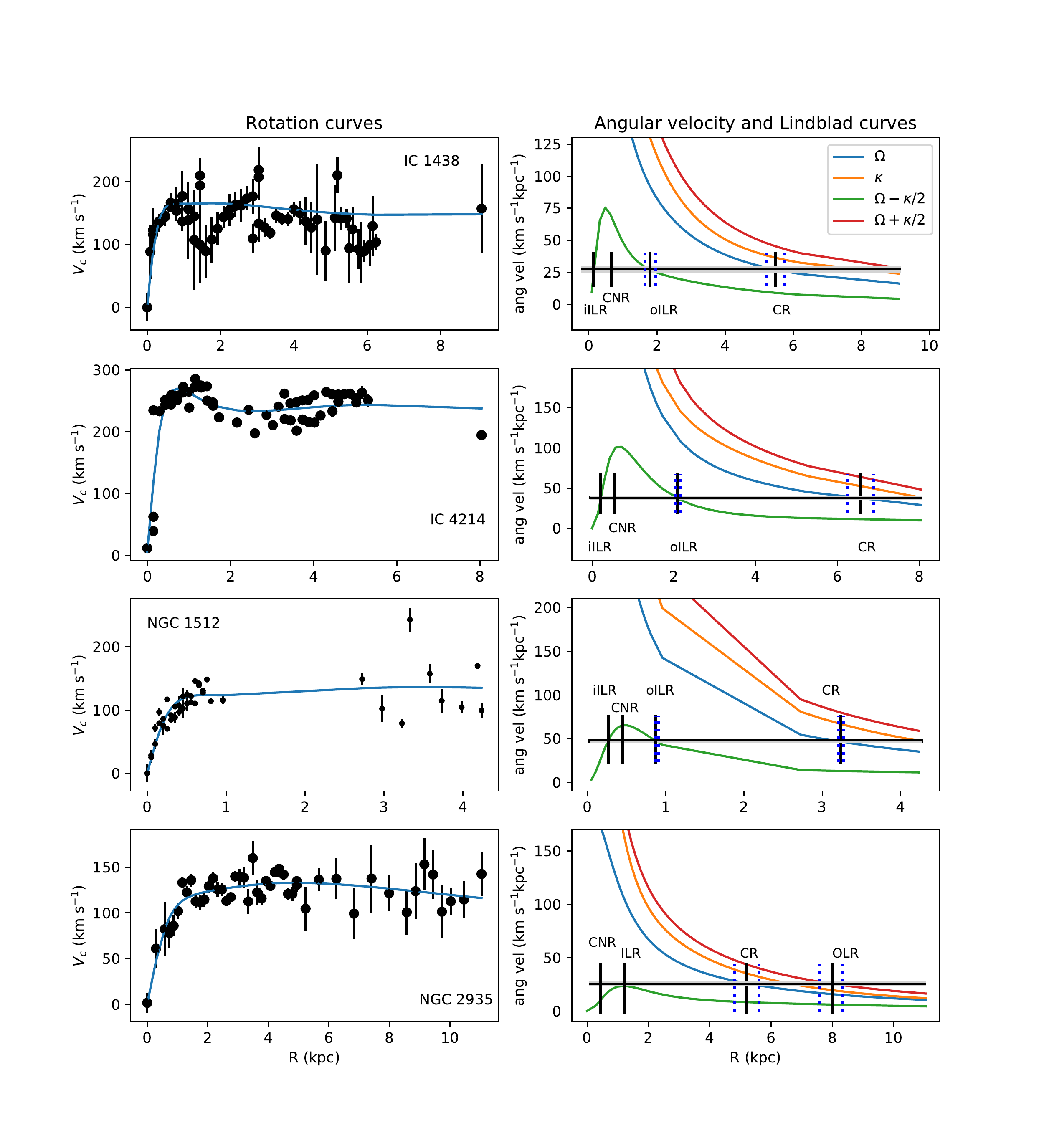}  
\caption{Left panels: rotation curves fitted with two or three components corresponding to axisymetric Miyamoto$-$Nagai potential. In all cases, V$_{circ}$ is 0 near the center. This is due to limited spatial resolution that makes the gravitational sphere of influence of the supermassive black hole not resolved in the rotation curves. Right panels: angular velocity curves and Lindblad curves, in units of km s$^{-1}$ kpc$^{-1}$. The horizontal black and gray solid lines indicate the perturbation pattern angular velocity and its uncertainty. The vertical blue dotted lines show the uncertainties of the position of the resonances.}
\label{fig:curves}
\end{figure*}
\addtocounter{figure}{-1}

\begin{figure*}[!ht]
\centering
\includegraphics[angle=0,trim = 28mm 0mm 0mm 0mm,clip, width=21cm]{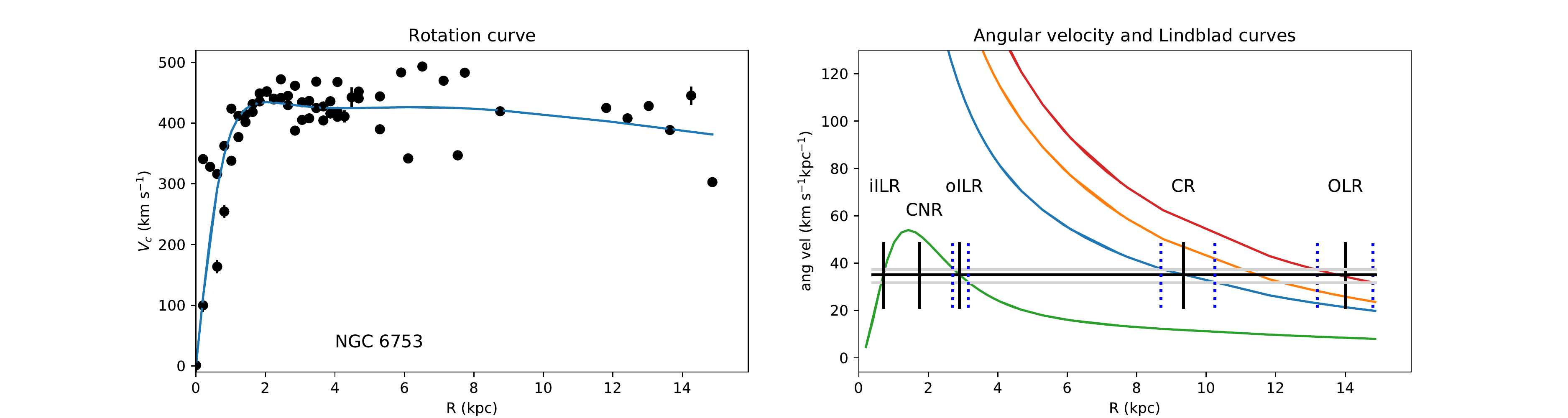}  
\caption{(continued)}
\end{figure*}

Table \ref{tab:reso} lists where the different resonances of the studied galaxies are located (in units of kpc), the pattern speed values, $\Omega_{p}$ (in units of km s$^{-1}$ kpc$^{-1}$), the radial lenght of the bar (in units of arcsec and kpc) and the dimensionless parameter $\mathcal{R}$.

\begin{table*}[!ht]
\caption{Angular velocity pattern and resonances}             
\label{tab:reso}      
\centering          
\begin{tabular}{c c c c c c c c }      
\hline\hline       
          
Galaxy       & $\Omega_{p}$              & ILR   & CR    & OLR  & r$_{bar}$ & R$_{bar}$ & $\mathcal{R}$         \\
             &  (km s$^{-1}$ kpc$^{-1}$) & (kpc) & (kpc) & (kpc)&  (arcsec) &  (kpc)    &                  \\        
    1         &    2                     &  3    &  4    &   5  &     6     &     7   &  8                      \\       
\hline                    
 IC 1438 & 27 $\pm$ 2 & iILR = 0.13 -- oILR = 1.8 $\pm$ 0.2 & 5.5 $\pm$ 0.3 &--& 20.6 $\pm$ 0.9 & 3.5 $\pm$ 0.1 & 1.6 $\pm$ 0.1  \\ 
 IC 4214   & 38 $\pm$ 2  & iILR = 0.21 -- oILR = 2.1 $\pm$ 0.1 & 6.6 $\pm$ 0.3 & --& 28.6 $\pm$ 0.9 & 4.1 $\pm$ 0.1& 1.6$\pm$ 0.1   \\
 NGC 1512  & 47 $\pm$ 1  & iILR = 0.26 -- oILR = 0.87$\pm$ 0.05 & 3.2 $\pm$ 0.1 &--& 60 $\pm$ 4 & 3.0 $\pm$ 0.2  & 1.1 $\pm$ 0.1  \\ 
 NGC 2935  & 26 $\pm$ 2  &  1.2 $\pm$ 0.2   & 5.2 $\pm$ 0.4  & 8.0 $\pm$ 0.4      & 24.3$\pm$ 0.7 & 3.3 $\pm$ 0.1 & 1.5 $\pm$ 0.2  \\ 
 
 NGC 6753  & 35 $\pm$ 3  & iILR = 0.72 -- oILR = 2.9 $\pm$ 0.2  & 9.4 $\pm$ 0.8  & 14.0 $\pm$ 0.8 & --  & --  & --   \\                 
\hline  
\end{tabular}
\footnote{Galaxy name (Col. 1), pattern angular velocity (Col. 2), position of the Inner Lindblad Resonance, ILR (Col. 3), location of the co-rotation Resonance, CR (Col. 4), position of the Outer Lindblad Resonance, OLR (Col. 5), radius of the bar in arcsec (Col. 6), radius of the bar in kpc (Col. 7), and the dimensionless parameter defined as $\mathcal{R}$= R$_{CR}$/R$_{bar}$ (Col. 8).}
\end{table*}

\subsection{Spatial profiles}\label{sec:profiles}

\begin{figure*}[ht!]
    \begin{center}
     \subfigure{\includegraphics[width=0.55\textwidth]{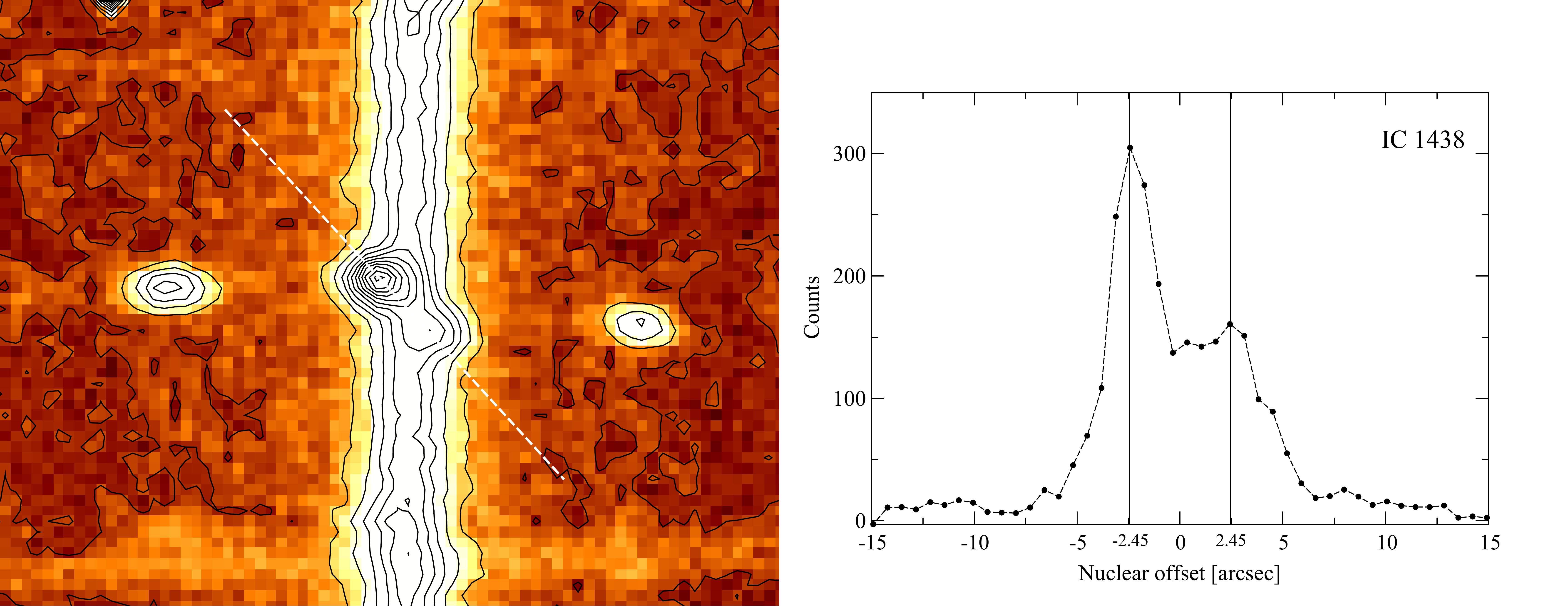}} 
    \subfigure{\includegraphics[width=0.55\textwidth]{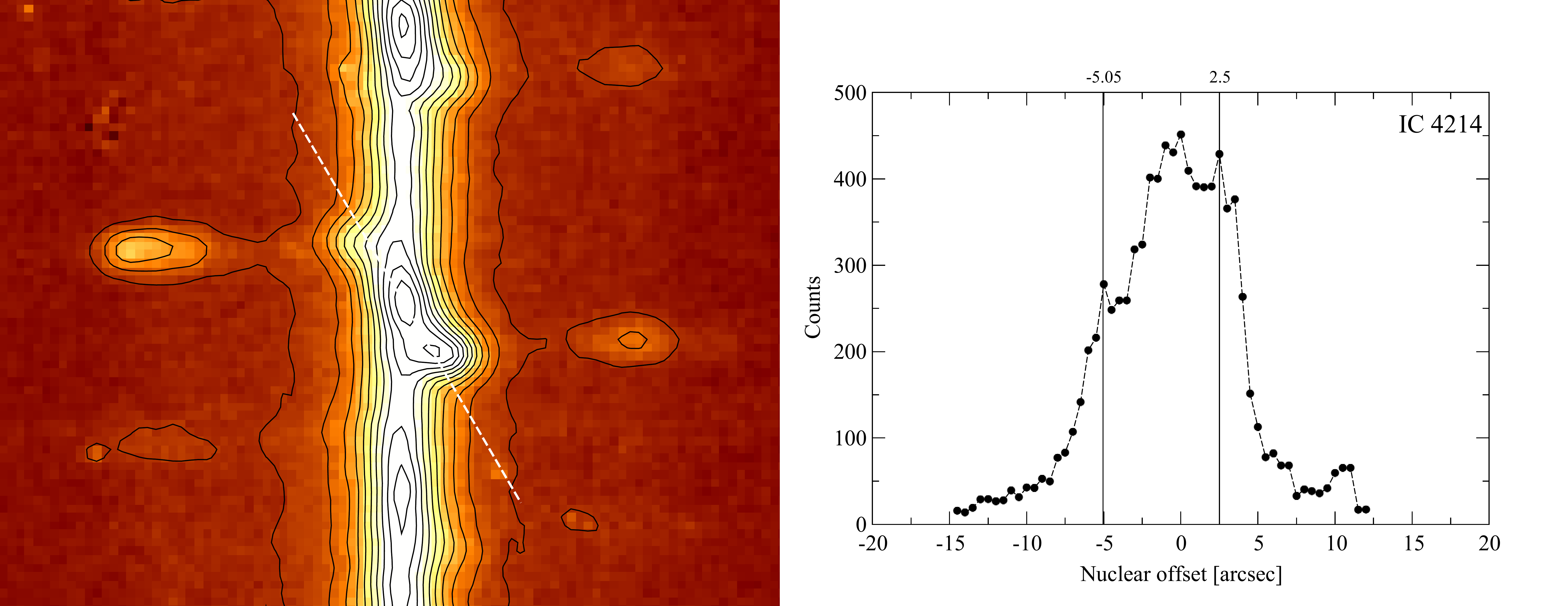}} 
    \subfigure{\includegraphics[width=0.55\textwidth]{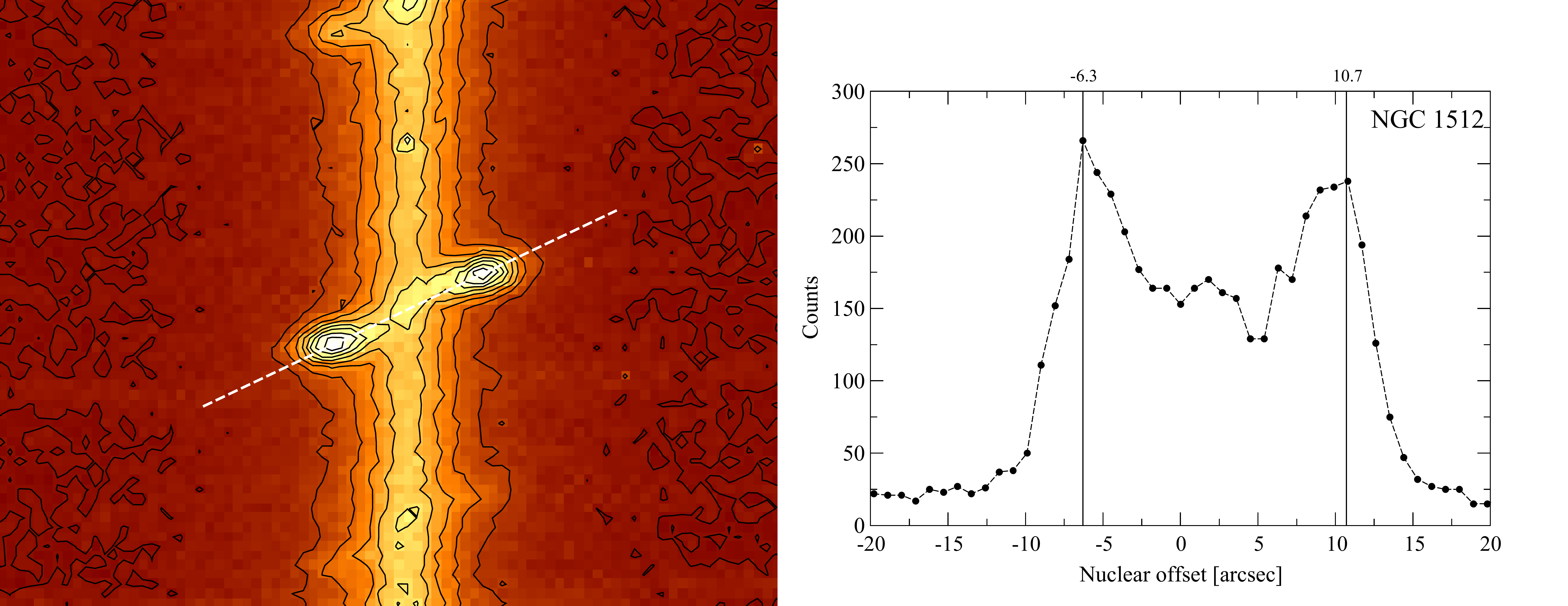}}
   \subfigure{\includegraphics[width=0.55\textwidth]{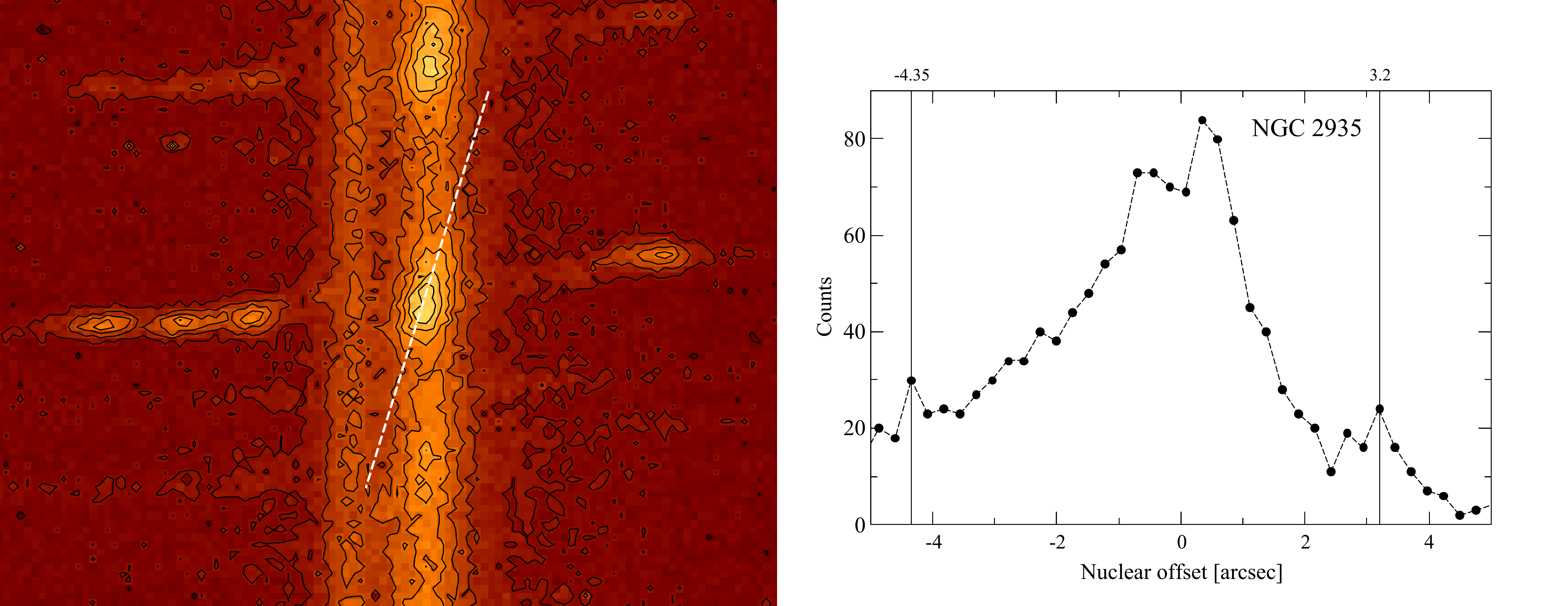}}
    \subfigure{\includegraphics[width=0.55\textwidth]{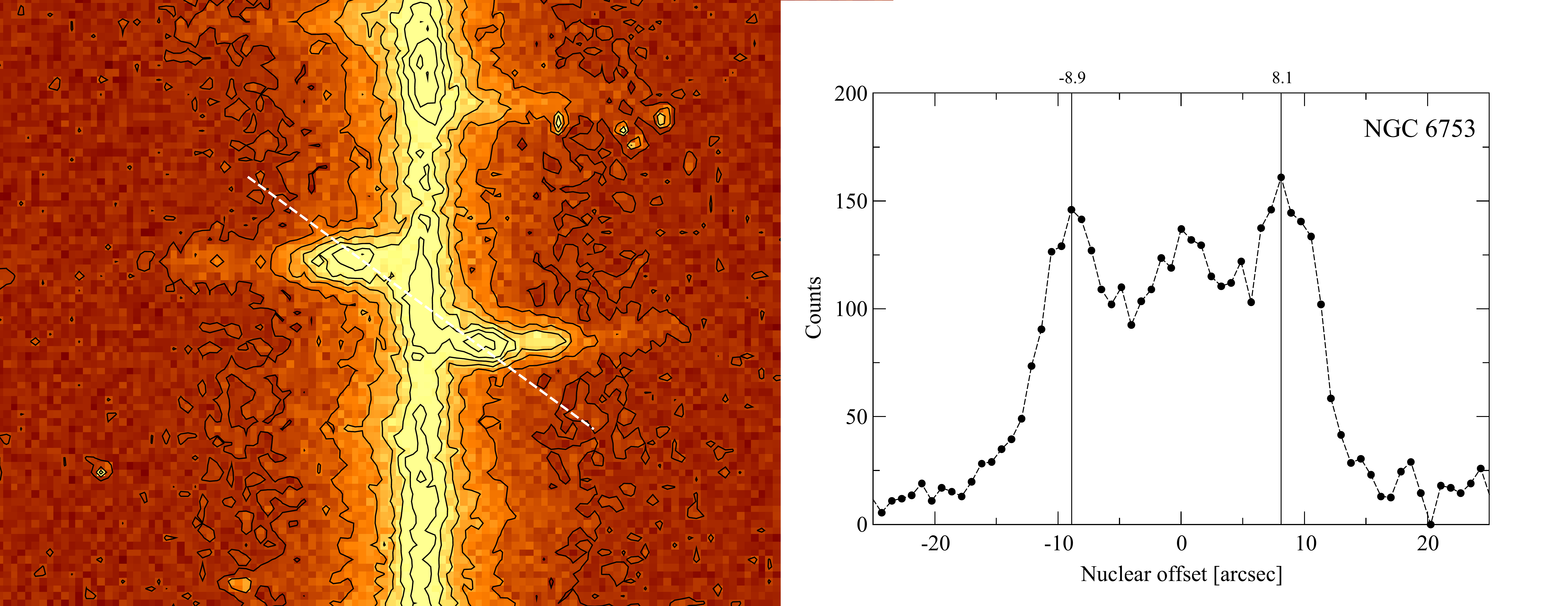}}
    \caption{\textit{Left:} zoom-in of the H$\alpha$ region in the 2D observed spectra. The horizontal direction is the spatial coordinate while the vertical direction represents the wavelength coordinate ascending up. The white dotted line indicates the direction of the extraction in order to construct the radial profiles. \textit{Right:} H$\alpha$ radial profiles extracted from the 2D spectra in an oblique direction, and projected on the spatial direction. Vertical lines indicate the  H$\alpha$ peak intensities at both sides of the galaxy center, that we associate with the CNR.}
    \label{fig:Profiles}
\end{center}    
\end{figure*}

To determine the CNR radius, we used the 2D spectra and generated the H$\alpha$ spatial emission radial profiles (Fig.\ref{fig:Profiles}). As can be seen in Figure \ref{fig:Profiles}, we perform an extraction by joining the emission peaks. This direction is oblique to both the spatial and spectral directions, so it is then necessary to project the profile over the spatial direction to have the distance in arcsec. In this way we are using the dispersive capacity of the diffraction grating to separate the emission and to be able to resolve and measure the rings radii. Unfortunately, we only have the information in the direction of the slit, so the PA of the ring is unknown. Although it is possible to measure the PA of the CNR from images \citep[e.g.,][]{Pogge1989, Buta1993, Comeron2010}, the discrepancy in both PA and radius between different authors shows that it is not a simple task. In addition, in the case of spectra it is easier to correctly isolate the H$\alpha$ emission  to define the ring. Another question that requires a proper definition would be which point of the CNR range is considered its radius. For the present work we decided to measure the distance between the emission peaks on both sides. The nucleus of the galaxy is defined as the continuum emission peak in the H$\alpha$ spectral region. In this way we avoid uncertainties due to diffuse emission in the internal or external region of the ring.

The CNR of the five galaxies studied in this paper have been previously measured in other works. As we will see below (Sec.\ref{sec:discussion}), these measurements vary from one work to another. Most of these previous studies use images in different bands to determine the CNR radius, obtaining distinct values. For completeness and looking for consistency in the measurement criteria, we have performed our own determinations of the CNR radio through the H$\alpha$ profiles (Figure \ref{fig:Profiles}). It should be noted that we have information only at the direction of the observed PA and this may differ from the actual PA of the CNR. In the comparison with the values from the  literature we carried out in Sec.\ref{sec:discussion}, these differences are considered.

\section{Discussion} \label{sec:discussion}

In this section we consider each galaxy separately, taking into account what was obtained in each one, and comparing in some cases with other results found in the literature. Then we will see how the obtained results fit or not within a particular scenario. Before getting into the analysis, it is important to make some considerations about the measured values.

The inclination angle of a galaxy, governed by the ratio of the semi-major and minor axis of the object, may vary with the distance to the center, as it may be the case of a warped disk \citep[e.g.,][]{Bosma1978,Briggs1990,Reshetnikov1999, Beckman2004,Saha2009, Buta2015}. In addition, the PA$_{kin}$ may also vary with radius of the galaxy and may be differ from the PA$_{phot}$. Taking all this into consideration, our assumption of a "perfect" disk may include some uncertainties which are very difficult to quantify. However, when comparing with values from the literature and, where possible, showing the effects of possible variations in the measured quantities, we can see that our approximations are valid and provide solid results within the reported uncertainties.  

Concerning the bar length, the measurement methods used in different works \citep{Athanassoula2002} can yield values that differ by as much as a factor 2 \citep{Comeron2010}. For this reason, in the present work we have visually measured the galaxy bar lengths \citep[e.g.,][]{Kormendy1979, Martin1995,Hoyle2011} directly from 2MASS images, since it is in these images in the near IR where the contribution of the more evolved stellar population is greater, as well as the contrast with the gaseous disk, which allows us to better visualize the extent of the bar.

Determining the position of the CNR and its radius is not an easy task either. In this work we have measured the radius of the rings, as we said before, from the H$\alpha$ profile. In the comparison with the values from the literature, our approach was to use our own measurement when the dispersion of values in the literature was very large. In all cases, the dispersion of measured values could give us an idea of the uncertainties involved in the determination of the CNR radius.

The uncertainties in the position of the resonances were determined by successive measurements of the CR (OLR in the case of NGC 6753) radii visually from DSS images and doing statistics. In this way we were able to take into account possible errors due to the determination of the center of the galaxy and asymmetries in the disk.

\subsection{Notes on individual objects}\label{sec:notesonobjects}

\textit{IC 1438}: According to \citet{Buta2009}, it is a weakly barred galaxy with a faint outer ring and presents enhanced star formation in nuclear and inner rings. Through isophotal fitting to DSS R-band images, we measured an inclination angle of 32$^{\circ}$, in agreement with previous results \citep[e.g.,][]{{Boker2008}}. The parameters of the fitted Miyamoto$-$Nagai components imply the presence of three galactic subsystems such as an intermediate bulge$-$disk, a disk, and an external halo, with a total mass of 2.4 $\times$ 10$^{11}$ M$_{\odot}$. We measured a bar radial length of 20\farcs6 $\pm$0\farcs9, equivalent to (3.5 $\pm$ 0.1) kpc, in agreement with the value obtained by \citet{Comeron2010}. We assumed that the CR is located at (5.5 $\pm$ 0.3) kpc from the center. Furthermore, this galaxy shows an angular velocity pattern of $\Omega_{p}=$ (27 $\pm$ 2) km s$^{-1}$ kpc$^{-1}$. Related to this, the dimensionless parameter $\mathcal{R}$ results 1.6 $\pm$ 0.1, consistent with a slow bar \citep[e.g.,][]{Debattista2000, Guo2019}. 
According to the H$\alpha$ profile (Figure \ref{fig:Profiles}), the CNR radius is 2\farcs45, equivalent to 0.41 kpc.
However, other values are found in the literature. For example, \citet{Boker2008} obtained a CNR radius of 0.537 kpc, with a PA$_{CNR}$ of 81$^{\circ}$. In addition, according to \cite{Buta1993} and \cite{Comeron2010}, the CNR semi-major axis is 0.650 kpc and 0.680 kpc, respectively. All these values show a dispersion of about 25\% and it may be due to the different PAs considered. Assuming even the most extreme possible values, it can be said that the CNR is located between the iILR (0.13 kpc) and the oILR (1.9 kpc). Our data does not allow us to have information about the location of the OLR, but it could be located at around 10 kpc from the center, considering the end of the spiral arms.  
On the other hand, our slit PA is 127$^{\circ}$, in agreement with the NED reported PA$_{phot}$. However, \cite{Boker2008} consider a PA of 145$^{\circ}$ for this galaxy. This difference of 18$^{\circ}$ in PA, would produce a negligible difference in the $\Omega_{p}$ value, and, as a consequence, in the location of the iILR, and a difference of 0.05 kpc in the location of the oILR.\\

\textit{IC 4214}: It is a weakly barred galaxy \citep{Moss1999} and according to the position of the galaxy in diagnostic diagrams based on line flux ratios, this object presents LINER activity \citep{Saravia2001}. These authors also report minimal star formation in the galaxy nucleus. Through isophotal fitting from DSS R-band images, we obtained an inclination angle of 54$^{\circ}$, which is consistent with the value published by NED.
The parameters of the fitted Miyamoto$-$Nagai components imply the presence of two galactic subsystems and a total mass of 1.6 $\times$ 10$^{11}$ M$_{\odot}$. We measured a radial length bar of R$_{bar}=$ 28\farcs6 $\pm$ 0\farcs9, equivalent to (4.1 $\pm$ 0.1) kpc, which is in agreement with the value obtained by \citet{Comeron2010}. We assumed that the CR is located at the end of the bar, at (6.6 $\pm$ 0.3) kpc from the center of the galaxy, with a dimensionless parameter $\mathcal{R}=$ 1.6 $\pm$ 0.1. Furthermore, this object presents an angular velocity pattern of $\Omega_{p}=$ (38 $\pm$ 2) km s$^{-1}$ kpc$^{-1}$. It can be seen in Figure \ref{fig:Profiles} that the CNR is located about 0.6 kpc (3\farcs8) from the nucleus. According to \citet{Buta1993}, the major axis of the ring is 0.146 arcmin, equivalent to a semi-major axis of 0.632 kpc (4\farcs38). \citet{Comeron2010} reports a ring semi-major axis of 7\farcs6, equivalent to 1.15 kpc. Despite the dispersion of reported values, and considering even the the extreme measurements, the CNR is located between the iILR (0.21 kpc) and the oILR (2.04 kpc). As in the previous case, we do not have any information about the location of the OLR. According to our measurements, the spiral arms seem to end at about 53$\farcs$7 ($\sim$ 7.75 kpc) from the center of the galaxy.\\

\textit{NGC 1512}: Through isophotal fitting from DSS R-band images, our measurements of the major and minor axis of the galaxy give an inclination angle of 54$^{\circ}$, in agreement with other works (see NED). The rotation curve fitting considering two component of Miyamoto$-$Nagai potential gives a total mass of 2.7 $\times$ 10$^{11}$ M$_{\odot}$. According to our measurements, the bar has a total length of 2 arcmin, equivalent to a radial length of 60" $\pm$ 4" and (3.0 $\pm$ 0.2) kpc. The CR was assumed to be at the end of the gap between the bar and the spiral arms, as mentioned before, at (3.2 $\pm$ 0.1) kpc from the galactic center. The system angular velocity pattern is $\Omega_{p}=$ (47 $\pm$ 1) km s$^{-1}$ kpc$^{-1}$. Related to this, the parameter $\mathcal{R}$ is 1.1 $\pm$ 0.1, consistent with a fast bar \citep{Debattista2000, Rautiainen2008, Guo2019}. According to \citet{Ma2017}, this galaxy has a nuclear ring with a total stellar mass of $\sim$10$^{7}$M$_{\odot}$ and a young average age of $\sim$ 40 Myr. Figure \ref{fig:Profiles} shows that the CNR has a radius of 8\farcs5, equivalent to 0.42 kpc. According to \citet{Buta1993}, the CNR has a major axis of 0.280 arcmin, equivalent to a semi-major axis of 8\farcs40 or 0.420 kpc. In addition, \citet{Comeron2010} report a CNR with a radius of 0.480 kpc. These values present a much lower dispersion than the previous ones and indicate that the CNR is also located between the iILR (at 0.26 kpc) and the oILR (at 0.90 kpc). The extension of the rotation curve does not allow us to estimate the location of the OLR. It is also very difficult to determine the end of the spiral arms, since from a radius of 3 arcmin, this galaxy presents very open arms, probably due to interactions, being possible to determine from DSS images an arm 8.5 arcmin from the center. This galaxy was also observed by the Hubble Space Telscope (HST) several times in the past. We used UV archive images in the F220W filter\footnote{Proposal ID: 4804, PI: Maoz, Dan.} to measure the CNR radius and to compare with our measurements. The ring measurement in the F220W filter images give R$_{CNR}=7\farcs7$ in acceptable agreement with our measurements, taking into account the spatial resolution and possible physical spatial differences between different bands.\\

\textit{NGC 2935}: This galaxy exhibit some lobesideness in its morphology, and the radial velocity curve is asymmetric (see Figure \ref{fig:radialvelocity}). \citet{Florido2012} observe non-resolved emission and they found knots of gas at large radii, corresponding to HII regions. According to our isophotal fitting from the DSS R-band image, this object has an inclination angle of 45$^{\circ}$, consistent with previous results (see NED). The rotation curve was fitted considering the S$-$E part of the radial velocity curve, mainly because of the irregularity that the N$-$W part presents. The radial velocities of descending values in the range of 1 $-$ 5 kpc (see Figure \ref{fig:radialvelocity}) is consistent with the detection of a strong incoming flux along the N-W semi-major axis of the bar. Although the inflow dominates the inner regions on the N-W side, there is also gas at higher radii with circular velocities with very similar values to those on the S-E side. Considering that, the fit of only the S-E side is perfectly representative of the circular velocities of the galaxy.

The two Miyamoto$-$Nagai fitted components under this assumption, gives a total mass of 4.3 $\times$ 10$^{10}$ M$_{\odot}$. This object presents a bar with a radial length of R$_{bar}=$24\farcs3 $\pm$ 0\farcs7, equivalent to (3.3 $\pm$ 0.1) kpc and we determined that the CR lies at (5.2 $\pm$ 0.4) kpc from the galactic nucleus. This yields an $\mathcal{R}=1.5$ $\pm$ 0.2, indicating that we are in the presence of a slow bar \citep[e.g.,][]{Rautiainen2008, Debattista2000, Guo2019}. The calculated $\Omega_{p}$ is (26 $\pm$ 2) km s$^{-1}$ kpc$^{-1}$ and according to the H$\alpha$ profile (Figure \ref{fig:Profiles}), the CNR has a radius of 0.52 kpc (3\farcs77), in agreement with \citet{Aguero2016}, who report a radius of 0.45 kpc (3\farcs26). Also, according to \citep{Comeron2010}, the CNR has a semi-major axis of 3\farcs6, equivalent to $\sim$0.50 kpc, and \citet{Buta1993} measured a CNR diameter of 0.106 arcmin, equivalent to a radius of 3\farcs18 and 0.44 kpc. The CNR is confined to a region between the center of the galaxy and the ILR, which is located at 1.2 kpc (see Figure \ref{fig:curves}). According to our estimation, the OLR is positioned at $\sim$ 8.3 kpc. The arms, from DSS images, end at 1.65 arcmin ($\sim$13.56 kpc), away from the OLR. This could be probably due to the high asymmetry commented above that disturb the galaxy disk. This object, however, presents a particular situation, taking into account what could generate a small variation in $\Omega_{p}$. On the one hand, there would be no ILR if the pattern speed were increased (see Figure \ref{fig:curves}). On the other hand, if $\Omega_{p}$ were decreased, there would be two ILRs. In this case, the CNR would be located inside the iILR.   \\

\textit{NGC 6753}: It is a grand-design spiral galaxy that shows a very pronounced spiral structure \citep{Windhorst2002}. From DSS R-band images, our estimation of the major and minor axis gives an inclination angle of 40$^{\circ}$, which is in agreement with previous results (see NED). The two Miyamoto$-$Nagai fitted components gives a total mass of 4.2 $\times$ 10$^{11}$ M$_{\odot}$. We visually inspect 2MASS images of this galaxy and we do not see any bar, in agreement with \citet{Comeron2010}. We estimated the OLR radius by assuming that the spiral arms end before the OLR. With this assumption the OLR is at (14.0 $\pm$ 0.8) kpc from the galactic nucleus. This way, the calculated $\Omega_{p}$ in this object is (35 $\pm$ 3) km s$^{-1}$ kpc$^{-1}$. According to our results, the corotation resonance is at (9.4 $\pm$ 0.8) kpc. This value is consistent with an obscured region in the vicinity where the spiral arms arise.  
According  to the H$\alpha$ profile (Figure \ref{fig:Profiles}), the CNR radius is 8\farcs50, equivalent to $\sim$ 1.89 kpc, in agreement with previous results \citep{Aguero2016, Comeron2010}. The higher value found in the literature corresponds to a CNR with a diameter of 0.35 arcmin, equivalent to a radius of 10\farcs5 or 2.16 kpc \citep{Buta1993}. Even considering the extreme values, the CNR is located between the iILR (at 0.7 kpc) and the oILR, which is at (2.9 $\pm$ 0.2) kpc.\\

\section{Summary and conclusion}
\label{sec:summary}

Throughout this work we have studied a sample of 5 spiral galaxies in order to determine the radii of the CNR and the location of the Lindblad resonances. To carry out this study, we have obtained and analyzed medium resolution spectra in the range 6200 \AA \ to 6900 \AA \ in order to measure the H$\alpha$ emission line.

Once the radial velocity curves (Figure \ref{fig:radialvelocity}) were determined, we have constructed the rotation curves, to which we fitted two or three components of an axisymmetric Miyamoto$-$Nagai potential. 
According to our results, the rotation curves of 4 of the 5 galaxies were fitted considering the presence of only two galactic subsystems, i.e. a central ellipsoidal component and a flattened disk. For these objects, it was not necessary to fit an external spherical component (halo). Considering the galaxy IC 1438, three components were fitted to the rotation curve, which are the central ellipsoidal, the flattened disk and the halo components.

Through the velocity curves fitting, we were able to construct the angular velocity curves and the Lindblad curves.
This way, we calculated the location of the Lindblad resonances.
In order to measure the circumnuclear radius, we used the H$\alpha$ emission spatial radial profiles obtained from the 2D observed spectra.
For NGC 1512, NGC 2935 and NGC 6753, the obtained values of the CNR radii are in agreement with previous results in the literature. For IC 1438 and IC 4214, the values found in this work show differences of 20\% and 10\%, respectively, comparing with previous results.
Taking into account all the measurements and estimations of the CNR radii, and even considering the extreme values, the CNR are located in all cases between the iILR and the oILR, or between the center and the ILR, as in the case of NGC 2935. This fact is in perfect agreement with previous results \citep[e.g.,][]{Combes1996, Buta1996, Sheth2000, Comeron2014, Li2015}. Related to this, as we mentioned in Section \ref{sec:notesonobjects}, NGC 2935 presents a particular case, considering what would generate a small change in $\Omega_{p}$. If the pattern speed were increased, it would result in no ILR. On the contrary, if $\Omega_{p}$ were decreased, there would be two ILRs, and the CNR would be located inside the iILR. There are, however, other galaxies studied in the literature which also present only one ILR  \citep[e.g.,][]{Combes1996, Shlosman1999, Comeron2014}.

In summary, although the location of the resonances is known, this information is insufficient to predict the exact radius of the CNR. Although there will be a range of radii where the CNR can be found, it is not enough to know the position of the resonance to know the exact location of the CNR, in agreement with the results of \citet{Li2015}. In this scenario, the fact that the CNRs are located inside the ILRs is not synonymous to say that the ring formation is due to the ILRs.  According to the analysis of \cite{Comeron2010}, galaxies with strong bars host smaller rings. In addition, these authors say that galaxies with weak bars can host both larger and smaller nuclear rings. This is in agreement with the results of \citet{WoongKim}, who use hydrodynamic simulations. Recently, \citet{Seo2019} found through three-dimensional simulations that rings are very small when they first form and grow in size over time. These authors found that the size of the CNR depends on the pattern speed, the central mass concentration and the bar strength. Related to this, \citet{Li2017} suggest that the ring size increases almost linearly with the mass of the central object (e.g., a black hole).

In addition, \citet{Li2015} find through simulations that CNR are in the range of $x_{2}$ orbits. Related to this, \citet{Regan2003} say that formation of CNR are more related to $x_{2}$ orbits than Lindblad resonances. However, according to some authors, the simulations of the last mentioned work do not have enough resolution to study in detail the structures of gas in the central regions of the galaxy \citep{Li2015}.

All the above considerations, however, are consistent with each other, taking into account that the size and extent of the $x_{2}$ orbital family correlate with the location of the inner resonances \citep[e.g.,][]{Contopoulos1989,Athanassoula1992b}. In this scenario, the $x_{2}$ orbits are in the region confined between the inner resonances. 

In this work we showed that different authors, using observations in different bands and with varied spatial and spectral resolutions, obtain similar results in relation with the position of CNR. We have shown that reducing uncertainties in measured parameters hardly brings new elements to the issue of CNR location. Our next step we will delve deeper into the study of the link between the presence of the CNR, the AGN duty cycle, and the life cycle of bars. Placing all these pieces in the same evolutionary scenario represents one of the current challenges in the study of galaxy evolution.

 
%

\begin{acknowledgements}
We want to thank Horacio Dottori and Walter Weidmann for fruitfull discussion. We also appreciate the helpful comments and suggestions made by the anonymous referee, which improved this article. This work was partially supported by the Consejo Nacional de Investigaciones Cient\'ificas y T\'ecnicas (CONICET, Argentina), Secretaría de Ciencia y Técnica
de la Universidad Nacional de Córdoba (SeCyT, Project Number: 33820180100080CB), and a GRFT2018 grant from MinCyT, C\'ordoba, Argentina.
Based on data acquired at Complejo Astronómico El Leoncito, operated under agreement between the Consejo Nacional de Investigaciones Cient\'{i}ficas y T\'ecnicas de la Rep\'ublica Argentina and the National Universities of La Plata, C\'ordoba, and San Juan.
This research has made use of the NASA/IPAC Extragalactic Database (NED) which is operated by the Jet Propulsion Laboratory, California Institute of Technology, under contract with the National Aeronautics and Space Administration.
We made an extensive use of the following python libraries: http://www.numpy.org/, http://www.scipy.org/, and http://www.matplotlib.org/.
This research has made use of SAO Image DS9, developed by Smithsonian Astrophysical Observatory.
\end{acknowledgements}

\bibliographystyle{yahapj}
\bibliography{Bibliography}

\end{document}